**Title**: Intra-protein binding peptide fragments have specific and intrinsic sequence patterns

**Authors**: Yuhong Wang*[1], Junzhou Huang[2], Wei Li[3], Sheng Wang[2], Chuanfan Ding [4]

**Affiliations**:

[1] School of Information Science and Engineering, Ningbo University, 818 Fenghua Road, Ningbo 315211, P. R. China

[2] Department of Computer Science and Engineering, The University of Texas at Arlington
Address: 500 UTA Blvd, ERB 640, Arlington, TX, 76019

[3] School of life science, Jilin University, 2699 Qianjin Street, Changchun 130012, P. R. China

[4] Department of chemistry and Laser Chemistry Institute, Fudan University, Shanghai 200433, P. R. China

**Abstract**: The key finding in the DNA double helix model is the specific pairing or binding between nucleotides A-T and C-G, and the pairing rules are the molecule basis of genetic code. Unfortunately, no such rules have been discovered for proteins. Here we show that similar rules and intrinsic sequence patterns between intra-protein binding peptide fragments do exist, and they can be extracted using a deep learning algorithm. Multi-millions of binding and non-binding peptide fragments from currently available protein X-ray structures are classified with an accuracy of up to 93%. This discovery has the potential in helping solve protein folding and protein-protein interaction problems, two open and fundamental problems in molecular biology.

**One Sentence Summary**: Classification of binding and non-binding intra-protein peptide fragments using feed-forward neural network

Introduction

Protein folding and protein-protein interaction are two fundamental, long-standing problems in molecular biology, and their importance can hardly be overestimated. The protein folding problem is to predict three dimension structure (3D) of a protein from its amino acid sequence (1D) (*1*). The protein-protein interaction (PPI) is to predict specific binding/interaction between two or more proteins (*2*). Life depends upon its components, these components' functioning, and information flow between them. Protein is one fundamental component of life, and its function depends upon 3D structure. PPI is the molecule basis of information flow.

Experimental approaches for determination of protein structure and PPI have advanced at an ever-faster rate (*2, 3*), but they remain expensive, time-consuming, and insufficient. For example, it is difficult to detect weak, but biological important interactions between proteins. While computational approaches are fast and inexpensive, their current roles remain supplementary. It remains impossible to predict protein structures and PPI *de novo* (*2, 3*) despite the huge advances in computing power.

This study started from our earlier interests in binding or spatially close peptide fragments in globular proteins (*4*). Current computational approaches for protein folding and PPI problem starts from the assumption that protein's native conformation corresponds to its global free energy minimum (*1*) and binding peptide fragments are brought together after 3D structures are formed. However, we did observe interesting patterns between intra-protein binding peptide fragments. Thus we proposed an alternative mechanism parallel to the DNA double helix model: binding peptide fragments are formed first and drive the formation of protein 3D structure and PPI. Unfortunately, available protein structure data in early 1990s was not sufficient for further exploration.

**Results**

In this study, we examined this alternative hypothesis, and our main thinking is that if this alternative hypothesis is true, binding peptide fragments must have specific and intrinsic sequence pattern that are distinct from non-binding ones. If sufficient number of samples is collected, binary classification algorithm in machine learning can be applied to test and identify such intrinsic patterns.

We focused upon specific interactions between two and three short peptide fragments, named as peptide triad (PT) and duo (PD), respectively, after common music terms. A binding peptide triad (BPT) and duo (BPD) and a non-binding peptide triad (NBPT) and duo (NBPD) are defined as PT and PD having all pair-wise distances between center residues < 5.0 Å and > 30 Å, respectively.

From 12,946 X-ray protein structures (*5*), we extracted 1.2-3.5 millions of BPTs, 1.4-4.5 millions of NBPTs, and 0.4-0.9 millions of BPD and NBPDs (Table 1). We designed a neural network (Figure 1) and performed supervised deep learning classification algorithm on the combined 2.6-8.0 and 0.8-1.9 millions of PT and PD samples, respectively. The input is the peptide triads or duos. Each hidden layer consists of 256 nodes. The output layer has two neurons for binding or non-binding.

The combined samples are randomly split into three data sets: 80% for training, 10% for validation, and 10% for test. The neural network was trained by minimizing the "cross-entropy" loss function using the ADAM optimizer (*6*), a mini-batch size of 128, and other optimized parameters (Table S1). The training process was monitored by checking accuracy of the validation data set, and terminated when no further improvement was observed. The trained models were applied to the test data set for benchmarking in terms of accuracy, area under the ROC curve (AUC-ROC), F-Score, precision and recall. For purpose of negative control, the labels (binding or non-binding) of samples were randomized, and the same training procedure and benchmarking were performed.

For PTs of 7 residues, the loss function of training data set dropped fast in the first 10 iterations, followed by much slower decrease (Figure 3). The prediction accuracy on validation data set increased fast in the first 10 iterations, followed by much slower improvement. For PTs of 7 residues of randomized labels, the loss function decreased fast in the first a few iterations and then stayed constant; the accuracy stayed at 0.5 throughout the training process. PTs of 2, 3, 4, 5, and 9 residues and PDs of 3, 5, 7, and 9 residues had very similar profiles.

The trained models were applied into test data sets and the performance benchmarks are listed in Table 1. For PTs of 2-9 residues, the accuracy increases from 0.74, 0.79, 0.84, 0.912, to 0.931 and then comes down to 0.923. PTs of 7 residues have the best accuracy of 0.931 and AUC-ROC of 0.979 (Figure 2), and this finding seems to be consistent with recent screening results (7). PDs of 3-9 residues have the accuracy of 0.620, 0.841, 0.836, and 0.770. PDs of 5 residues have best accuracy of 0.841 and AUC-ROC of 0.911. No meaningful models could be learned from PTs and PDs of randomized labels (Table 3), and the AUC-ROC have perfectly random values of 0.5.

## Discussion

The up to 93% of accuracy (Table 1) and AUC-ROC of 0.979 (Figure 2) from multi-millions of PT and PD samples shows that intra-protein binding peptide fragments do have specific and intrinsic sequence patterns. The learned patterns, encoded in the neural network model, are unlikely computational artifacts. First, no models could be learned from negative control or PTs and PDs of randomized labels. Second, substantial changes in the neural network structure, including number of hidden layers and nodes, and training parameters will not significantly affect the classification performance. Third, no difference exists in amino acid composition between BPTs/BPDs and NBPTs/NBPDs (Figure 5, pvalue = 1.0).

PTs have a much better performance than PDs. The best accuracy for PDs is 0.841, significantly lower than the best one for PTs (0.931). This difference is unlikely due to size of the input layer of the neural network. We achieved accuracy of 0.80 with 6.9 millions of PTs of three residues and the input layer size of 180 (3x3x20). In contrast, for 0.8 million samples of PDs of 9 residues with the input layer size of 360 (2x9x20), the accuracy is 0.770. Thermodynamically, three peptide fragments should have stronger binding than two, and this provides a further evidence for the intrinsic nature of the learned neural network models.

In this study, all BPTs and BPDs from known protein X-ray structures are predicted with accuracies of up to 93% and 84%, respectively, and they are apparently helpful in predicting topology and large-scale structure of proteins from amino acid sequence (8). BPTs plus BPDs are likely important force in forming large scale structures of proteins, and they may provide an answer to the Levinthal's paradox (9). For a protein of 150 residues, assuming a minimum amino acid separation of 10 between two binding fragments, we have roughly 15 chunks. The possible combinations of choosing 3 out of 15 is 455. Thus, a protein, itself a computing machine, may not need to search through $3^{150}$ possible conformations to find global free energy minimum.

Machine learning algorithms have been applied into prediction of protein contact map with various degree of success (10). These efforts are based upon the assumption that two residues of a protein are brought together and in contact after the protein's 3D structure is formed; thus they use entire protein sequences in the machine learning algorithm. This study is based upon the alternative hypothesis that specific binding between short peptides are the driving force. The excellent performance of the trained neural network supports this alternative hypothesis, and it also benefits from the much larger data sets for training and testing.

**Materials and Methods**

Protein structure data

We used 12,946 protein X-ray structures from Protein Data Bank (PDB) (*5*) to extract intra-protein peptide fragments, either binding or non-binding. These proteins are from the precompiled culled PDB list (*11*), and the goal of the list is to create a non-redundant coverage for all available protein structures. Proteins in this list have an amino acid percent identify < 50%, a resolution better than 2.0 A, and a R-factor smaller than 2.5.

Extraction of peptide triads and peptide duos

An intra-protein BPT is defined as three peptide fragments of a protein having all three pair-wise distances between center residues smaller than 5.0 Å. A NBPT is defined as three peptide fragments having all three pair-wise distances between center residues greater than 30 Å. Choosing 30 Å is to produce a balanced training data set, and smaller cutoffs do not affect the training results. To avoid redundancy, if the positions on the amino acid sequences of all three fragments of two PTs, either binding or non-binding, are less than 9 residues away from each other, these two PTs are considered as the same, and only one PT is used. Duplicated PTs (about 5-10%) were eliminated, and the numbers of unique BPTs and NBPTs of 2-9 residues, extracted from 12,946 protein database entries, are given in Table 1.

An intra-protein BPD is similarly defined as BPT, and the difference is in the number of fragments (three vs two). One BPT essentially consists of three BPDs. To learn the model for two peptide fragments only, BPDs from BPTs are excluded in the training and test of PDs. We also performed training and testing with BPDs including BPTs, and observed no significant differences.

To perform the deep neural network training, each amino acid is encoded by 20 bit vector or 20 neurons. For PT of seven residues, for example, the total size of the input vector or number of neurons in the input layer is 3x7x20 = 480. Among the 480 bits or neurons, only 21 (3x7) have 1s, and all rest 0s.

Deep learning

Deep Learning (*13*) methods, as representation learning methods, allow deep neural networks discovering the representations from raw data for specific tasks such as classification and detection. Supervised learning is the most common form of machine learning which deep learning improves the state-of-the-art of most supervised learning problems. With the help of the ground truth or label of data set, deep learning can learn better representation to predict such ground truth. A loss function captures the distance between the current output of the neural network and the ground truth, then the network propagates the error backwards to adjust all the parameters (weights) in the neural network. In this way, the loss or distance can be significantly reduced after the training process. The binding and non-binding peptide fragments classification is supervised learning with the ground truth as if the peptide fragments are binding or non-binding. Thus, we use deep learning to learn better features and get better classification performance.

We designed a fully connected feedforward neural network of one input layer, four hidden layers, and one output layer for binding and non-binding classification (Figure 1). For PTs of 2, 3, 4, 5, 7, and 9 residues, the input layer consists of 120, 180, 240, 300, 480, and 540 nodes, respectively. Each hidden layer consists of 256 nodes or neurons. In each hidden layer, the fully-connected layer is followed by the activation function of Rectified Linear Units (ReLU) (14) which can introduce nonlinearity into the presentation learning. After the hidden layers, Softmax layer is used as the classification layer (or the output layer of two nodes for binding or non-binding). Significant changes in the neural network, including number of hidden layers and nodes, will not significantly affect the classification performance, and 4 hidden layers of 256 nodes tend to produce good results. Backpropagation is used for training the network (15).

The input to the *j*th node of a hidden layer is calculated according to following equation, where $w_{i,j}$ is the weight connecting *i*th node of previous layer and $\theta_j$ is the bias.

$$X_j = \sum_i w_{i,j} + \theta_j$$

All hidden layers use the Rectified Linear Unit as the activation function, and output layer uses Softmax function as the activation function.

We used "cross-entropy" with L2 regularization as the loss function according to the following equation:

$$H = \sum_i \sum_j y'_j \log(y_j) + \lambda \sum_i (w_i^2 + \theta_i^2)$$

Where *i* denotes *i* th training sample, *j j* th class, *y* is the predicted probability distribution, *y'* is the true distribution (the one-hot representation of the label), and $\lambda$ is the coefficient for L2 regularization.

Optimization of the loss function is carried out by mini-batch of a size 128 and the ADAM optimizer (6), which is implemented as tf.train.AdamOptimizer in the Tensorflow library (www.tensorflow.org). The regularization coefficient and starting learning rate were optimized after a grid search (Table 2).

The neural network training and prediction were performed on CyberpowerPC SLC2400C desktop with Intel core i7 and 8GB Nvidia GeForce GTX 1080 graphic processing unit, installed with Ubutun distribution of 16.10, python 3.4, CUDA driver version 8.0, cuDANN version 5.1, and Tensorflow 0.11rc. The python program was written to implement the neural network model (Figure 1) and optimize the loss function.

BPTs/BPDs and NBPTs/NBPDs were randomly split into three data sets: 80% for training, 10% for validation, and 10% for test (Table 1). The training process was constantly monitored by checking the accuracy of the validation data set, and it was terminated in about 3000 epochs and about 20 hours when either no further improvement was observed or the improvement was deemed too slow to be meaningful. The trained models were applied to the test data set for benchmarking.

For negative control, the label of each PT and PD was randomly assigned as binding (1) or non-binding (0), and the same training procedure and benchmarking were performed.

Training process

The loss of training data set of for peptide triads of 7 residues is plotted versus iterations in Figure 3. The plots for other training data sets are very similar. For peptide triads of 7 residues, after a dramatic drop in the first 10 iterations, the loss keeps decreasing, but at a significantly reduced speed. We stopped the training process after about 3000 iterations. It is interesting to see a relatively quick reduction in loss function between iteration 128 and 256. For peptide triads with random labels, no noticeable reduction is observed after first 10 iterations.

The prediction accuracy of validation data set shows similar profiles (Figure 4). For peptide triads of 7 residues, the accuracy has a fast increase in the first 10 iterations. Afterward, it keeps increasing, but at a much reduced speed. Corresponding to fast decrease in loss function between iteration 128 and 256, we also see a relatively quick increase in accuracy. For peptide triads of randomized labels, the accuracy stays at 0.5 throughout the training process.

Amino acid composition

We also compared amino acid composition difference between binding and non-binding peptide triads and observed no difference (pvalue = 1.0, Figure 5).

**Figure legends:**

Figure 1. Illustration of the forward deep learning model for classification of binding and non-binding peptide fragments.

Figure 2. ROC curves for binding peptide triads of 7 residues. Total sample size of test data is 330,305. The AUC-ROC are 0.979 and 0.500 for test data of correct and randomized labels, respectively.

Figure 3. Model loss on training data sets for peptide triads of 7 residues with correct and randomized labels. Sample size of the training data set is `2,644,109`.

Figure 4. Model accuracy on validation data sets for peptide triads of 7 residues with correct and randomized labels. Sample size of the validation data set is `330,869`.

Figure 5. Amino acid composition of binding and non-binding peptide triads of 7 residues. A student t test gives a pvalue of 1.0.

**Figures**

Figure 1

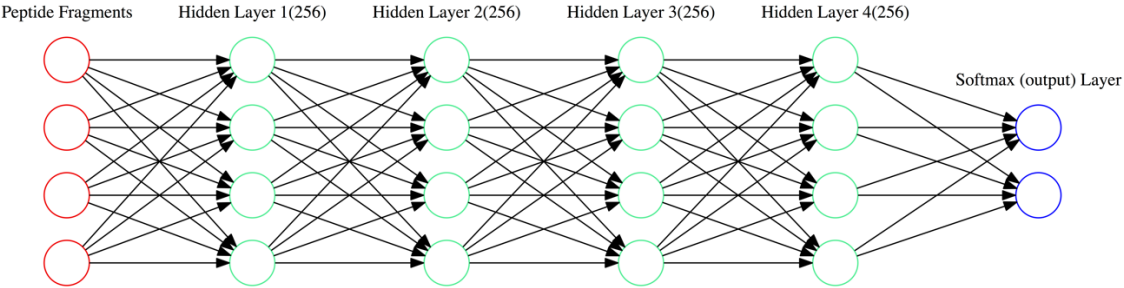

Figure 2

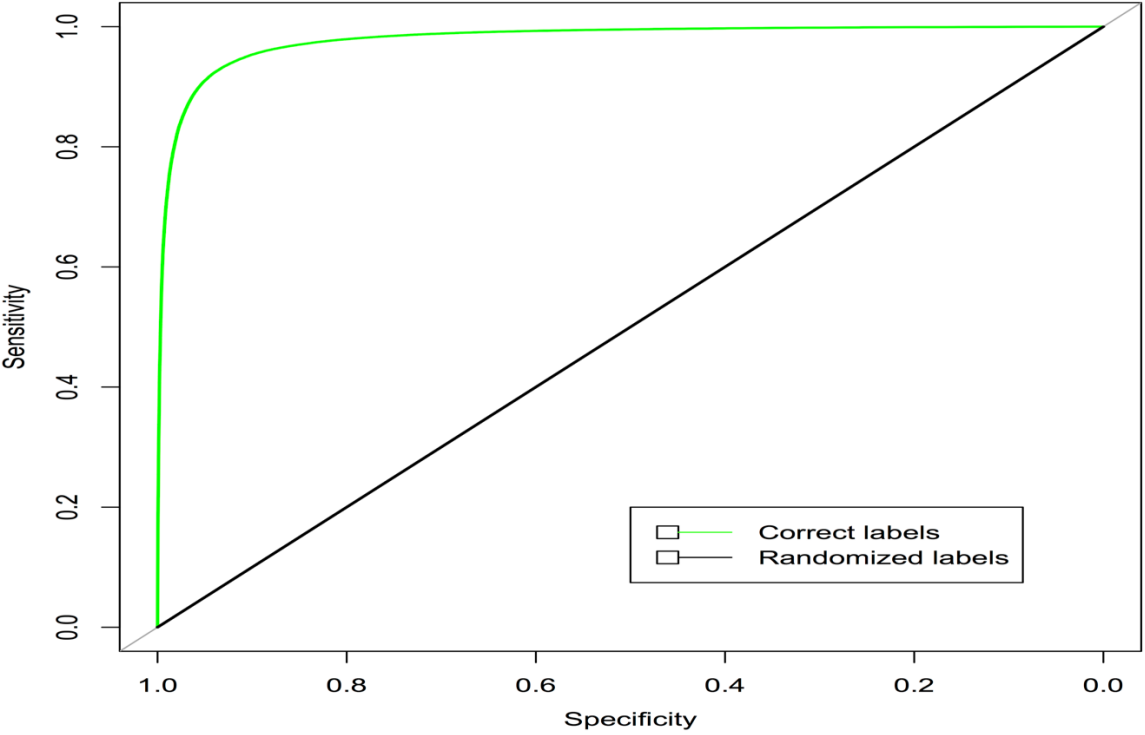

Figure 3

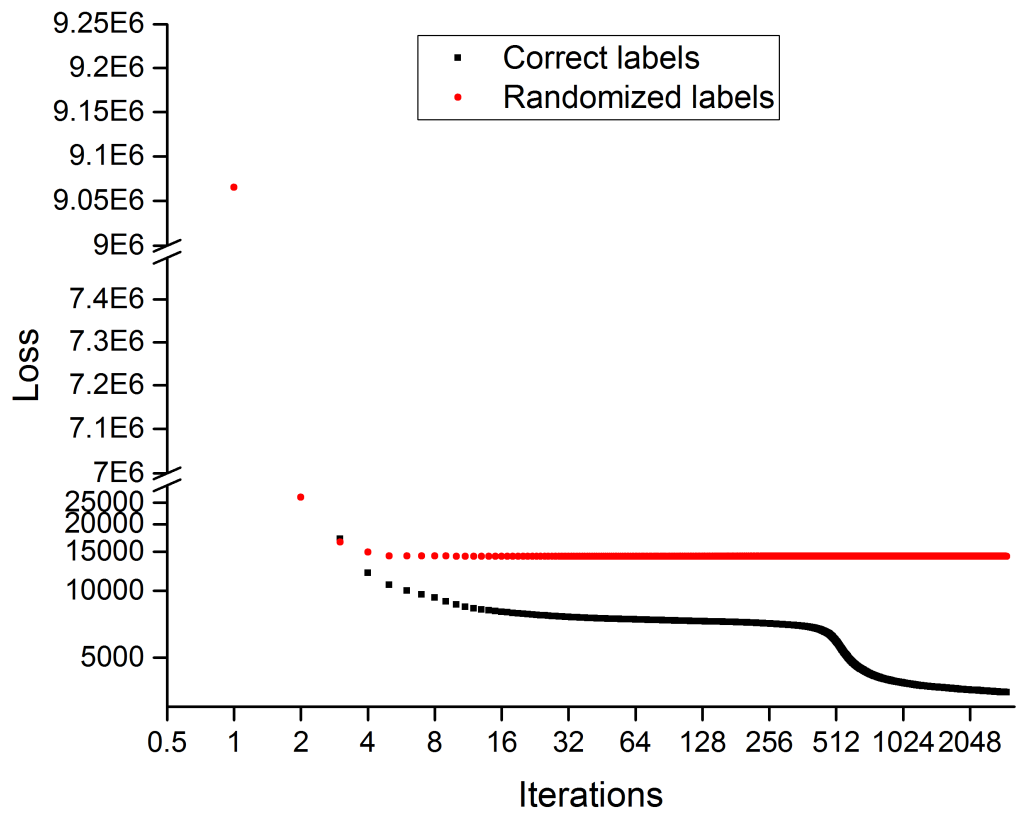

Figure 4

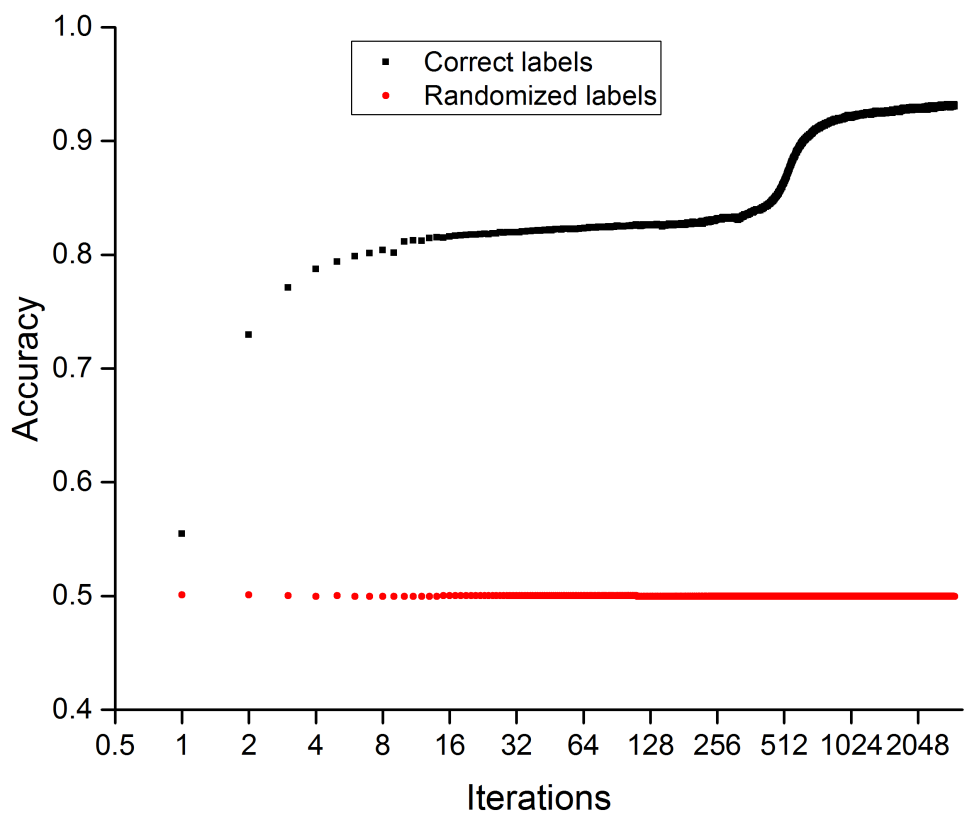

Figure 5

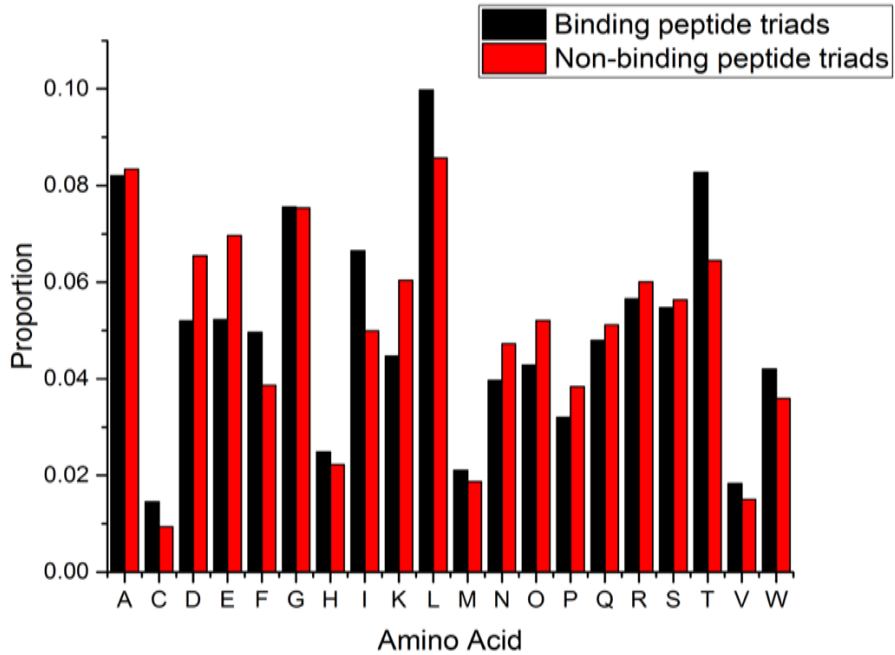

**Tables**

Table 1. Classification results of binding and non-binding peptide fragments on test data sets in terms of accuracy, area under the ROC curve (AUC-ROC), F-Score, precision, and recall. Loss function is optimized using the ADAM optimizer a mini-batch size is 128. Other optimized parameters are given in Table S1. [1] Number of binding peptide fragments samples, and [2] number of non-binding fragments samples.

| Peptide fragments | No of BPFS [1] | No of NBPFS [2] | Final loss | Accuracy | AUC-ROC | F-Score | Precision | Recall |
|---|---|---|---|---|---|---|---|---|
| Peptide Triads | | | | | | | | |
| 3x2 | 3,506,094 | 4,573,534 | 26,417 | 0.739 | 0.807 | 0.675 | 0.735 | 0.624 |
| 3x3 | 3,202,563 | 3,727,467 | 19,058 | 0.793 | 0.871 | 0.768 | 0.797 | 0.742 |
| 3x4 | 2,454,016 | 2,821,820 | 11,256 | 0.849 | 0.924 | 0.834 | 0.852 | 0.817 |
| 3x5 | 1,943,073 | 2,346,432 | 5,491 | 0.915 | 0.969 | 0.906 | 0.911 | 0.902 |
| 3x7 | 1,561,153 | 1,744,130 | 3,450 | 0.931 | 0.979 | 0.927 | 0.938 | 0.917 |
| 3x9 | 1,276,502 | 1,398,539 | 2,815 | 0.923 | 0.975 | 0.919 | 0.915 | 0.923 |
| Peptide Duos | | | | | | | | |
| 2x3 | 938,992 | 972,945 | 7,463 | 0.620 | 0.668 | 0.577 | 0.639 | 0.526 |
| 2x5 | 692,955 | 658,592 | 2,689 | 0.841 | 0.911 | 0.841 | 0.861 | 0.822 |
| 2x7 | 526,614 | 506,836 | 1,572 | 0.836 | 0.905 | 0.835 | 0.856 | 0.816 |
| 2x9 | 419,945 | 420,781 | 1,238 | 0.770 | 0.845 | 0.758 | 0.800 | 0.721 |

Table 2. The regularization coefficient and starting learning rate for the neural network working; both were optimized after a grid search.

| Peptide fragments | Regularization coefficient | Starting learning rate |
|---|---|---|
| 3x2 | 0.0000010 | 0.0006 |
| 3x3 | 0.0000010 | 0.0006 |
| 3x4 | 0.0000010 | 0.0006 |
| 3x5 | 0.0000010 | 0.0006 |
| 3x7 | 0.0000025 | 0.0006 |
| 3x9 | 0.0000025 | 0.0008 |
| 2x3 | 0.0000010 | 0.0008 |
| 2x5 | 0.0000010 | 0.0008 |
| 2x7 | 0.0000010 | 0.0008 |
| 2x9 | 0.0000010 | 0.0008 |

Table 3. Classification results of binding and non-binding peptide fragments with randomized labels on test data sets in terms of accuracy, area under the ROC curve (AUC-ROC), F-Score, precision, and recall. Loss function is optimized using the ADAM optimizer a mini-batch size of 128. [1] Number of binding peptide fragments samples, and [2] number of non-binding fragments samples. Calculated probability in the binding output node for all test data set with randomized labels is constant and slightly below 0.5 [3] or above 0.5 [4].

| Peptide Fragments | No of BFSs [1] | No of NBFSs [2] | Final loss | Accuracy | AUC-ROC | F-Score | Precision | Recall |
|---|---|---|---|---|---|---|---|---|
| Peptide Triads | | | | | | | | |
| 3x2 | 3,506,094 | 4,573,534 | 34,992 | 0.501 | 0.502 | 0.668[3] | 0.501 | 1.000 |
| 3x3 | 3,202,563 | 3,727,467 | 30,027 | 0.499 | 0.499 | 0.666 | 0.499 | 1.000 |
| 3x4 | 2,454,016 | 2,821,820 | 22,864 | 0.500 | 0.499 | 0.000[4] | 0.000 | 0.000 |
| 3x5 | 1,943,073 | 2,346,432 | 18,585 | 0.502 | 0.500 | 0.668 | 0.502 | 1.000 |
| 3x7 | 1,561,153 | 1,744,130 | 14,319 | 0.500 | 0.500 | 0.667 | 0.500 | 1.000 |
| 3x9 | 1,276,502 | 1,398,539 | 11,586 | 0.500 | 0.500 | 0.666 | 0.500 | 1.000 |
| Peptide Duos | | | | | | | | |
| 2x3 | 938,992 | 972,945 | 8,284 | 0.501 | 0.500 | 0.000 | 0.000 | 0.000 |
| 2x5 | 692,955 | 658,592 | 5,857 | 0.498 | 0.500 | 0.665 | 0.498 | 1.000 |
| 2x7 | 526,614 | 506,836 | 3,625 | 0.503 | 0.504 | 0.523 | 0.503 | 0.544 |
| 2x9 | 419,945 | 420,781 | 2,519 | 0.501 | 0.500 | 0.505 | 0.500 | 0.510 |